# Winds from Luminous Late-Type Stars: II. Broadband Frequency Distribution of Alfvén Waves

V. Airapetian (CUA/NASA/GSFC), K. G. Carpenter (NASA/GSFC) and

L. Ofman (CUA/NASA/GSFC)

**Abstract**

We present the numerical simulations of winds from evolved giant stars using a fully non-linear, time dependent 2.5-dimensional magnetohydrodynamic (MHD) code. This study extends our previous fully non-linear MHD wind simulations to include a broadband frequency spectrum of Alfvén waves that drive winds from red giant stars. We calculated four Alfvén wind models that cover the whole range of Alfvén wave frequency spectrum to characterize the role of freely propagated and reflected Alfvén waves in the gravitationally stratified atmosphere of a late-type giant star. Our simulations demonstrate that, unlike linear Alfvén wave-driven wind models, a stellar wind model based on plasma acceleration due to broadband non-linear Alfvén waves, can consistently reproduce the wide range of observed radial velocity profiles of the winds, their terminal velocities and the observed mass loss rates. Comparison of the calculated mass loss rates with the empirically determined mass loss rate for α Tau suggests an anisotropic and time-dependent nature of stellar winds from evolved giants.

## 1. Introduction

At the latest stages of stellar evolution sun-like stars enter the red giant evolutionary phase losing a substantial fraction of their mass in the form of a relatively dense ($\geq 10^{-11}$ $M_{sun}$/yr) and slow-moving ($\leq 100$ km/s) wind. Large-scale convective motions of giants and supergiant stars deliver CN processed plasma from stellar interiors to the surface, where the wind injects it into interstellar medium, where it serves as a source for the formation of new stars and planets. However, despite its fundamental importance and decades of extensive theoretical and observational studies, the physical processes which drive massive and slow winds remain unknown.

Recent high-resolution spectroscopic observations have provided important constraints on the physical mechanisms of stellar winds. These observations have revealed four effects which are of critical importance in understanding the atmospheric dynamics. First, detailed examination of chromospheric emission lines from Fe II, O I and Mg II spectra shows clear evidence for the initiation and acceleration of the wind. For giant stars the wind appears to be initiated near the top of the chromosphere (Carpenter, Robinson & Judge, 1995). This constraint rules out the mechanism of radiative pressure on dust. The mechanism is not capable of lifting atmospheric plasma in red giants, because most dust is observed too far away from the stellar photosphere (Bester et al. 1996; Danchi et al. 1994). Second, there is evidence for strong turbulence within the chromospheres of both giant and supergiant stars. The C II] $\lambda 2325$ line is formed at temperatures between 5,000 and 10,000 K and is optically thin. It has an intrinsically narrow profile, so the line width primarily reflects the Doppler broadening caused by atmospheric turbulence. The deduced turbulent velocities range from 24 km/s for the K5 giant $\alpha$ Tau to 28 km/s for the K5 hybrid giant $\gamma$ Dra to about 30 km/s for the M3.4 giant $\gamma$ Cru. The fact that the chromospheric sound speed in these stars is typically less than 10 km/s implies that this turbulence is highly supersonic. This constraint suggests that sound waves driven by convection and/or pulsation do not contribute significantly to the initiation of the stellar wind (e. g. Castor 1981; Cuntz 1990; Judge & Cuntz 1993). While this mechanism can be important in depositing energy in the lower parts of the atmospheres of giants and supergiants (for instance, Lobel & Dupree 2001), it also fails to reproduce observed terminal velocities and mass loss rates from late-type giants (Sutmann and Cuntz 1995). Third, observations suggests that winds from cool evolved stars are at least two orders of magnitude cooler than that observed from the Sun and coronal giants (Harper 2001). This suggests that the thermal conduction cannot play any important role in driving winds in their atmospheres. Fourth, *FUSE* and *HST/STIS* observations of red giants have recently revealed that "virtually all late-type giants were hybrids to some extent, at least in terms of the $10_5$ K emissions" (Ayres 2005; Dupree et al. 2005). Observations of hot and broad (>100 km/s) UV lines suggests that the effects of magnetic fields can be a crucial factor in dissipating energy in the stellar chromosphere and depositing momentum to the stellar wind (Ayres et al. 2003).

The most promising mechanism to drive massive and slow winds from K and M giant and supergiant stars involves incompressible transverse magnetohydrodynamic (MHD) waves (or Alfvén waves) that propagate along the magnetic field in the stellar atmosphere (Osterbrock 1961; Hartmann & McGregor 1980; Hartman & Avrett 1984; Charbonneau & MacGregor 1995; Airapetian et al. 2000). Dissipation lengths of Alfvén waves are much larger compared with the compressive acoustic or MHD waves. Early models of Alfvén-driven winds from late-type giants (Hartman & MacGregor 1980) suggested that in order to drive massive winds at mass loss rates $\sim 10^{-7}$ $M_{sun}$/yr, the total energy flux in Alfvén waves, $F_A = \frac{1}{2}\rho_0 <\delta V_w^2> V_{A0}$, generated at the top of the chromosphere should be in the order of $\sim 10^5$ ergs/cm$^2$/s, where $<\delta V_w^2>$ is the mean square wave amplitude, $\rho_0$ and $V_{A0}$ are the density and the local Alfvén speed at the wind base. This implies that if the wind base density is $\rho_0 \sim 10^9 \, cm^{-3}$ (Robinson & Carpenter 1998), then the magnetic fields in the atmospheres of late-type giants should be in the order of ~1-10 Gauss if the area filling factor was included (Harper et al. 2009) .

This estimate of the magnetic field strength appears also to be consistent with the atmospheric magnetic fields required to confine the OVI emitting plasma observed in cool giants (Ayres et al. 2003; Dupree et al. 2005; Airapetian & Carpenter 2009). Moreover, GHRS/HST observations of non-thermal broadening in Fe II in the spectra of a number of red giants (Robinson et al. 1998; Carpenter and Robinson 1999) provide evidence for the existence of substantial supersonic turbulence (~ 24-27 km/s) distributed primarily along the radial and/or tangential directions in wind forming regions. The equipartition between the time-averaged (over the wave period) kinetic and magnetic energies of Alfvén wave fluctuations, $\rho_0 <\delta V_w^2> = \frac{<\delta B_w^2>}{4\pi}$, yields a magnetic field of ~ 1 Gauss. Recent direct observations of surface magnetic fields in a number of giants imply the presence of magnetic fields of a few Gauss (Tarasova 2002; Konstantinova-Antova et al. 2009). Another indication of the existence of magnetic fields in cool giants comes from VLBI observations of polarized SiO maser emission in evolved stars (Kemball et al. 2009) and observations of anisotropic mass losses in from a number of evolved stars as revealed by radio interferometric measurements of radial velocities of SiO, $H_2O$, and OH masers in AGB stars (Hofmann et al. 2000; Thompson et al. 2002).

Previous models of steady state, spherically symmetric winds driven by Alfvén waves suggested that in order to produce the low terminal velocities (that are less than half of the surface escape velocities) and the high mass loss rates, most of the energy and momentum should be deposited by Alfvén waves below the sonic point, while momentum addition beyond the sonic point will produce the fast and tenuous solar-like winds (Hartmann & McGregor 1980; Holzer & McGregor 1985). However, many important questions such as how waves propagate in the gravitationally stratified stellar atmosphere and damp their energy were left unspecified because they assumed an idealized Wenzel-Kramers-Brillouin (WKB) approximation. These studies parameterized the wave damping and adjusted its level to fit the observations. Later studies, including those of An et al. (1990), Velli (1993), Rosner et al. (1995), McGregor & Charbonneau (1995) and Charbonneau & McGregor (1995) extended the analysis to non-WKB waves and found that waves can be reflected from atmospheric regions where the gradient of the Alfvén velocity exceeds the wave frequency and that most of the momentum can be deposited below the sonic point. These studies used a linear treatment to infer the momentum deposition by Alfvén waves to the atmospheric plasma to estimate wind acceleration. The location of the sonic point for steady isothermal winds can be expressed as a function of the $\alpha$ parameter as $r_s = \frac{\alpha}{2} R_*$, where $R_*$ is the stellar radius, and dimensionless parameter $\alpha = \frac{R_*}{H}$, characterizes the ratio of the stellar radius to the pressure scale height at the wind base, $H = \frac{kT_0}{\mu m_H g}$, $g$ is the gravitational acceleration at the wind base, $\mu$ is the mean atomic weight of 1.67, and $T_0$ is the temperature of a gravitationally stratified spherically symmetric hydrostatic atmosphere. This parameters is typically greater than 10 for atmospheres of late-type luminous stars.

To examine the possibility of producing massive and slow winds by reflected Alfvén waves below the sonic point, Airapetian et al. (2000) computed the model of stellar winds driven by non-linear single-frequency Alfvén waves launched at the wind base into the gravitationally stratified atmospheres of giants and supergiants using a 2.5D fully non-linear resistive MHD code. Our simulations suggested that non-linear effects limit the length of the energy dissipation in subsonic regions and provide efficient momentum deposition through the non-linear coupling between the azimuthal and radial components of the velocity and magnetic fields in the first 3-5 stellar radii. This limits the magnitude of the terminal velocity of the simulated wind. To account for very large mass loss rates of $\sim 10^{-6} M_{sun}/yr$ that correspond to the wind from α Ori, the model required the magnetic field at the wind base, $B_0$ = 240 G. While such strong surface magnetic fields were thought to be unrealistically large, recent MHD dynamo models of late-type supergiants suggests the possibility that magnetic fields of comparable magnitude could be generated in their atmospheres (Freytag et al. 2002; Dorch 2004). Recently, Suzuki (2007) has applied a 1.5D MHD code to simulate winds as well as explain heating rates in the atmospheres of evolved giants across the "Linsky-Haisch dividing line" (Linsky & Haisch 1979). His model consistently treats acoustic and MHD wave energy dissipation in a fully ionized plasma assuming that waves are directly generated at the stellar photosphere by convective motions with the magnetic field of ~240 Gauss. However, the photospheric plasma in K-M giants is weakly ionized. This suggests due to ion-neutral coupling the damping lengths of Alfvén waves are about 4 orders of magnitude lower than that expected in fully ionized atmosphere, as shown by Khodachenko et al. (2004), and, therefore, Alfvén waves will be efficiently dissipated well below the top of the stellar chromosphere where the wind acceleration gets initiated. Moreover, because the pressure scale height in the cool photospheric layers is at least 4 orders of magnitude smaller than the characteristic wavelength of Alfvén waves (periods of a day and longer), upward propagating Alfvén waves will also suffer significant reflection from the regions of Alfvén velocity gradients. Indeed, if Alfvén wave speed jumps discontinuously as waves propagate from the stellar photosphere at the Alfvén speed, $V_{A,ph} = 0.7 \text{ km/s} (B_{ph} = 10 \text{ Gs})$, to the top of the stellar chromosphere at $r = 3R_*$ along the radially diverging open magnetic field at the local Alfvén speed, $V_{A0} = 138 \text{ km/s} (B_{ph} = 2 \text{ Gs})$, they will suffer the reflection with the coefficient, $R = \dfrac{V_{A0} - V_{A,ph}}{V_{A0} + V_{A,ph}} \sim 0.99$ (Davila 1991; Ofman 2002). This estimate suggests that only about 1% of the wave flux will leak out to the chromosphere. Therefore, the combined effect of the wave dissipation that reduces the wave flux by 4 orders of magnitude and predominant (99%) reflection of waves in a weakly ionized and highly stratified plasma of a red giant suggests that the Alfvén wave energy flux at the top of the chromosphere should be reduced by 6 orders of magnitude with respect to the wave flux generated in the photosphere. From the energy flux requirements to drive stellar winds discussed earlier, we conclude that the total energy flux at the photosphere should be on the order of

$F = \dfrac{1}{2} \rho_0 <\delta V_\phi^2> V_{A,ph} \sim 10^{11}$ ergs/cm$^2$/s. For the typical photospheric density and convective velocity in a red giant, this requires the Alfvén velocity ~1500 km/s, and, therefore, the existence of the photospheric magnetic fields of 10 kG! Meantime, the energy flux of photospheric turbulent motions that excite sound waves in the stellar atmosphere of a red giant is

$$P_{conv} = \dfrac{1}{2} \rho_{ph} V_{conv}^2 V_s,$$

where $\rho_{ph}$ is the photospheric density in a red giant according to MARCS models of Gustafsson et al. (2003), $V_{conv}$ is the convective velocity. Assuming $\rho_{ph} = 3.2 \times 10^{-8}$ gr cm$^{-3}$ according to MARCS models of Gustafsson et al. (2003) and $V_{conv} = 2$ km/s (Robinson et al. 1998) in the photosphere of $\alpha$ Tau, we obtain the maximum kinetic energy flux injected by convective motions $\sim 6 \times 10^{8}$ ergs/cm$^2$/s. This implies that if the compressible waves are converted into incompressible (Alfven) waves in the stellar chromosphere with the factor of ~0.1%, this can provide the energy flux of Alfven waves required to drive stellar winds. Therefore, we assume that Alfvén waves that drive stellar winds need to be generated at the top of the chromosphere by acoustic shocks produced by sound waves formed due to photospheric convection and/or pulsation.

In this paper we extend our stellar wind models driven by Alfvén waves generated at the top of the chromosphere to perform a parametric study of the effects of broadband Alfvén wave frequency spectra on the dynamics of stellar winds, such as terminal velocity and the mass-loss rate. Specifically, we develop four Alfvén wave-driven wind models that differ by the lowest frequency in the Alfvén wave spectrum at the wind base and study the trapping efficiency of waves and its effect on mass-loss rates and wind speed. We apply our models to K-M type giants stars and discuss their predictions.

The paper is organized as follows. In Section 2 we present the model equations and boundary conditions. In Section 3 we describe the model parameters for the simulated stellar atmosphere. The results of our MHD simulations as they apply to the wind from red giants are presented in Section 4. Section 5 discusses the implications of the wind models and describes some future improvements for wind models.

## 2. Time-Dependent 2.5 D MHD Simulations of Winds from α Tau -like Star

Here we apply a single fluid MHD approximation to treat the kinematics of stellar winds extending from the wind base to 20 stellar radii. The validity of this approximation is supported by HST/GHRS and STIS observations of relatively dense, $n_e \sim 10^9$ cm$^{-3}$, warm, $T_e \sim (1-3) \times 10^4$ K, slow moving ( V < 100 km/s) stellar wind plasma from K-M giants at the wind base (Robinson et al. 1998).

We use a time-dependent single fluid non-relativistic fully non-linear, resistive 2.5D MHD code in spherical geometry to simulate the propagation of Alfvén waves in the stellar atmosphere. We assume that a broadband spectrum of Alfvén waves are generated at the wind base and launched along a radially diverging magnetic filed within a specified cone in the gravitationally stratified stellar atmosphere of a red giant star. To study the dynamics of wind plasma we solve the normalized mass conservation equation (Equation 1), momentum equation (Equation 2) and induction equation (Equation 3) for the polytropic equation of state (Equation 4) presented as the following

$$\frac{\partial \rho}{\partial t} + \vec{\nabla} \bullet (\rho \vec{V}) = 0 \qquad (1)$$

$$\frac{\partial \vec{V}}{\partial t} + (\vec{V} \bullet \vec{\nabla})\vec{V} = -\frac{E_U}{\rho}\nabla p - \frac{1}{F_r r^2} + \frac{\vec{J} \times \vec{B}}{\rho} \qquad (2)$$

$$\frac{\partial \vec{B}}{\partial t} = \vec{\nabla} \times (\vec{V} \times \vec{B}) + S^{-1}\nabla^2 \vec{B} \qquad (3)$$

$$\left(\frac{\partial}{\partial t} + \vec{V} \bullet \vec{\nabla}\right)\frac{p}{\rho^\gamma} = 0 \qquad (4)$$

where the variables have been normalized to the stellar parameters, i.e., radial distance, $r = r/R_*$, is expressed in units of the stellar radius, $R_*$, time, $t = t/\tau_A$, in units of the Alfvén transit time over the stellar radius, $\tau_A = R_*/V_{A0}$, where $V_{A0} = \dfrac{B_0}{\sqrt{4\pi\rho_0}}$ is the Alfvén speed, $B_0$ and $\rho_0$ are the magnetic field and the plasma density at the wind base, $r = 1$. The plasma velocity, $\vec{V} = \vec{V}/V_{A0}$, magnetic field, $\vec{B} = \vec{B}/B_0$, plasma density, $\rho = \rho/\rho_0$, and plasma pressure, $p = p/p_0$, are normalized to $V_{A0}$, $B_0$, $\rho_0$ and the electron plasma pressure, $p_0$, at the wind base, respectively. The MHD equations (1-4) are written in dimensionless units and contain the following three dimensionless parameters: the Froude number is the square of the ratio of the Alfvén velocity to the escape velocity at the wind base, $V_{esc} = \sqrt{\dfrac{2GM_*}{R_{ch}}}$, $G$ is the gravitational constant, $R_{ch}$ is the height of the stellar chromosphere, $Fr = \dfrac{V_{A0}^2}{V_{esc}^2}$, the Euler number, is the square of the ratio of the sound speed to the Alfvén speed at the wind base, $Eu = \dfrac{p_0}{\rho_0 V_{A0}^2} = \left(\dfrac{c_{S0}}{V_{A0}}\right)^2$, where $c_{S0}$ is the local sound speed, and the Lundquist number, $S = \dfrac{4\pi V_{A0} R_{star}}{\nu c^2}$, is the ratio of the resistive diffusion time to the Alfvén transit time and measures the importance of plasma conductivity, where $\nu$ is the classical Spitzer resistivity of plasma parallel to the background magnetic field. The maximum value of the Lundquist number is limited by resolution-dependent numerical diffusion in the model. For the grid resolutions used in the simulations presented below we simulate the resistive plasma wind at the assumed uniform resistivity corresponding to $S = 10^4$.

The method of solving continuity, momentum and induction equations in fully ionized isothermal plasma has been described in detail by Ofman & Davila (1997). The solutions use the 4th order Runge-Kutta solver in time with upwind corrections, and 4th order derivatives in space. We have also applied fourth order smoothing (i.e., numerical viscosity) to stabilize the solution. The time step is determined by the Courant-Friedrichs-Lewy (CFL) condition.

The equations (1)-(4) are solved in 2.5 dimensions assuming azimuthal symmetry (i.e., $\partial/\partial\varphi = 0$, two spatial dimensions and three components of velocity and magnetic field). In our calculations we assume a uniform, radially diverging background magnetic field to be purely radial, the unipolar form

$$\vec{B}_0 = B_0 \vec{e}_r / r^2, \qquad (5)$$

within a cone-shaped atmospheric region with an opening angle of $45^0$ (see Figure 1). The radial density profile of the initially hydrostatic spherically symmetric isothermal atmosphere is

$$\rho_e = e^{\alpha(1/r-1)} \qquad (6)$$

In this study we focus on the wind dynamics in red giants driven by Alfvén waves on top of Parker flow with no specification of heating or cooling mechanisms in the energy equation. We assume that the wind plasma is nearly isothermal, i.e., $\gamma = 1.01$, in the equation of state (Equation 4). The boundary conditions at the wind base, r=1, are incoming characteristics approximated by zero-order approximation

$$B_r(1,\theta) = B_0, \ V_\theta(1,\theta) = B_\theta(1,\theta) = 0, \text{ and } V_r(1,\theta) \geq 0.$$

At the outer boundary, $R = 20R_*$ and at $\theta = \theta_{max}$ we use open boundary conditions with the symmetry imposed at $\theta = 90^0$.

In the present study we generalize our previous wind model (Airapetian et al. 2000) by introducing the Alfvén wave broadband frequency driver at the lower boundary (instead of a single frequency driver) as

$$B_\phi(t,\theta,r=1) = -V_d/V_{A0} \sum_{i=1}^{N} a_i \sin(\omega_i t + \Gamma_i(\theta)) \quad (7)$$

where $a_i = i^{-1/2}$, $\omega_i = \omega_1 + (i-1)\Delta\omega$, $\Delta\omega = (\omega_N - \omega_1)/N - 1$, and $\Gamma_i(\theta)$ in the phase that is the random function in space and time, $V_d$ is the Alfvén wave initial amplitude, $V_{A0}$ is the local Alfvén velocity at the wind base, $\omega_N = 5\omega_1$ is the highest frequency in the wave spectrum, and $N$=100 is the number of modes (Ofman 2004). The wind base in the computational domain is divided into 5 regions that generate the Alfvén waves with different $\Gamma_i(\theta)$.

**3. Model Parameters**

Our MHD simulations of the stellar atmospheric dynamics were performed for an axisymmetric single fluid MHD flow in spherical coordinates ($r,\theta,\phi$) in the meridional plane corresponding to $\phi = 0^0$ assuming an initially radially diverging (monopole) magnetic field in the gravitationally stratified stellar atmosphere. The calculations were conducted on the two dimensional r - θ grid of 800 uniformly spaced points in radial direction between $R = R_*$ (the wind base) and $20R_*$ and 400 uniformly spaced points in θ direction covering a cone-shaped atmospheric region with an opening angle of $45^0$.

In this study we simulated the wind from a typical α Tau-like cool giant star, with the observational and model parameters presented in Table 1, where the stellar mass, $M$ is expressed in units of $M_{sun}$, the stellar radius, $R_*$ is in units of $R_{sun}$, the plasma density, $n_0$, is in cm$^{-3}$, the sound speed, $V_s$, and the Alfvén speed, $V_A$, are defined at the wind base and expressed in km/s. The models of stellar winds were initialized with an isothermal Parker's (1960) wind flow at T=30,000K.

**Table 1. Summary of the observational and model parameters for the simulated α Tau like star**

| Parameter | Initial Value |
|---|---|
| *Stellar mass, $M_* / M_{sun}$* | 1.3[1] |
| *Stellar radius, $M_* / M_{sun}$* | 44[1] |
| *Mass loss rate ( $M_* / M_{sun}$ / yr)* | 1.6 x 10$^{-11}$[1] |
| *Wind terminal velocity, $V_\infty$ (km/s)* | 30[1] |
| *Mean turbulent velocity, $V_t$ (km/s)* | 24[1] |
| *Magnetic field at the wind base, $B_0$* | 2 G |
| *Temperature at the wind base, $T_0$* | 30,000K |
| *Parameter $\alpha$* | 11.238 |
| *Plasma density at the wind base, $n_0$* | $10^9$ cm$^{-3}$ |
| *Alfvén speed at the wind base, $V_A$* | 137.99 km/s |
| *Sound speed at the wind base, $V_s$* | 22.25 km/s |
| *Alfvén transit time, $\tau_A$* | 2.58 days |

Reference: (1) Robinson et al. (1998)

In the current simulations we use the observational parameters of a typical red giant, $\alpha$ Tau, as presented in Table 1. This results in $\alpha$ = 11.23 at $T_0$ =30,000K, and the sonic point, $r_s$ = 6.62 $R_*$.

For the given parameters presented in Table 1, the magnetic pressure in the stellar atmosphere completely dominates the thermal pressure at the wind base, so plasma motions are directed along the radial magnetic field. The radial profile of the Alfvén velocity for an isothermal stellar atmosphere, with the magnetic field and density profiles given by (5) and (6) respectively, in spherical geometry is

$$V_A = \frac{V_{A0}}{r^2} e^{\frac{\alpha}{2}(1-1/r)} \qquad (8)$$

The left panel of Figure 1 shows that the Alfvén speed increases exponentially up to $r = \frac{\alpha}{4} R_*$ or about first $3R_*$, and then decreases as $\sim \frac{1}{r^2}$ to infinity. The right panel of the figure shows that the gradient of the Alfven speed reaches the maximum at heights $\sim 2R_*$. An et al. (1990) studied the propagation of Alfvén waves in stratified atmospheres characterized by $\alpha$ parameter as an initial value problem and demonstrated that the wave reflection becomes significant at locations of large gradient of Alfvén velocity. Later, Velli (1993) expanded their model by considering the solutions of the stationary problem and studied how the transmission coefficient depends on wave frequency. Their analysis suggested that Alfvén waves at frequencies, $\omega \sim \omega_c$, should be trapped within the region specified by the maximum of the gradient of Alfvén speed. The right panel of Figure 1 shows that in the simulated atmosphere of a cool giant Alfvén waves should experience reflection at frequencies less than the critical frequency, $\omega_c = 3.94\tau_A^{-1}$. For the stellar parameters specified in Table 1, the critical frequency is ~17.7µHz with the corresponding wave period of 4.14 days.

To investigate the effect of the Alfvén wave frequencies on the wind terminal velocity and mass loss rates, we constructed 4 wind models driven by broad-band Alfvén waves further referred to as Models A, B, C and D with the lowest frequencies (and the corresponding wave periods) in the simulated broad-band spectrum of Alfvén waves presented in Table 2. In Models A, B, C and D we assumed the following lowest frequencies of the broad-band Alfvén wave spectrum, $\omega_1 = 1\tau_A^{-1}, 3\tau_A^{-1}, 9\tau_A^{-1}$ and $18\tau_A^{-1}$, or the corresponding wave periods of 16.33d, 5.44d, 1.8d and 0.9d respectively. This allows us to study the effect of partially reflected waves (Models A and B) and freely propagating waves (Models C & D) on the kinematic characteristics of stellar winds.

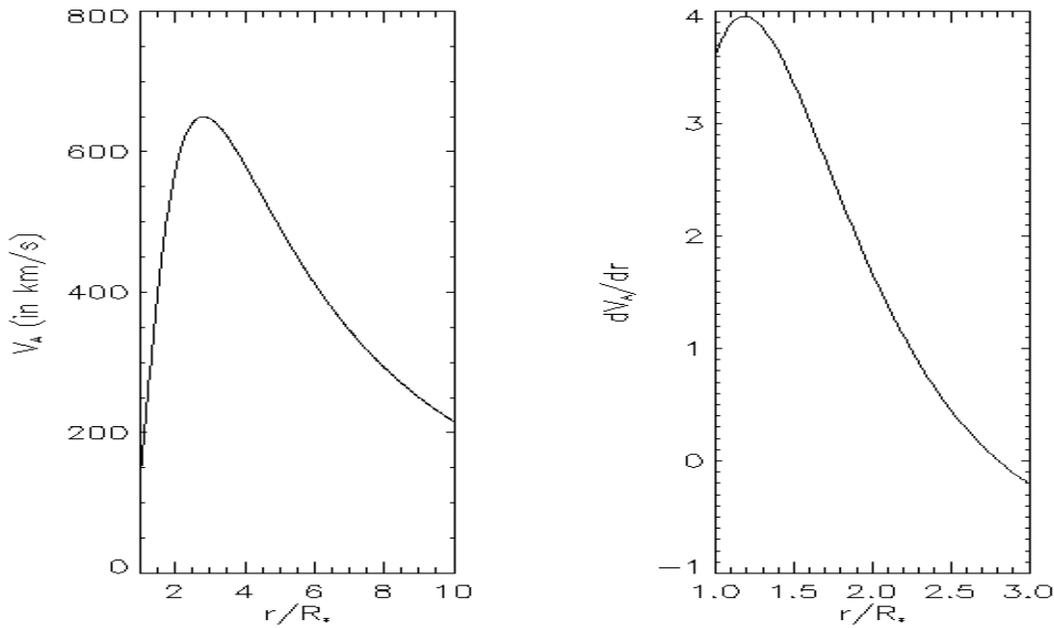

**Figure 1.** Initial radial profiles of Alfvén velocity (left panel) and the gradient of Alfvén velocity (right panel) in the simulated atmosphere of a red giant with α=11.23.

**Table 2.** Wave parameters of wind models

| Model Type | $\omega_1$ (in $\tau_A^{-1}$) | Wave period, P (days) |
|---|---|---|
| Model A | 1 | 16.2 |
| Model B | 3 | 5.4 |
| Model C | 9 | 1.8 |
| Model D | 18 | 0.9 |

The choice of the range of the lowest frequencies, $\omega_1$, in the simulated Alfvén wave spectrum comes from the assumption that Alfvén waves are generated by direct perturbations of magnetic fields at the wind base driven due to transverse velocity perturbations driven by non-linear sound waves propagating upward from the stellar photosphere into the chromosphere. The characteristic frequency of the sound wave driver can be estimated from the characteristic turnover time of a convective cell . The linear scale of a cell can be estimated from the scaling of cells obtained from 3D global hydrodynamic simulations of red giants by Freytag et al. (1997) as

$$l_{conv} = 2.5 \times 10^{-3} \frac{R_*}{R_{sun}} \frac{T_{eff,*}}{T_{ef,sun}} \frac{M_{sun}}{M_*} R_* , \qquad (9)$$

where $T_{eff,*}$ and $T_{eff,sun}$ are the effective temperatures of the star and the Sun respectively.

For convective cell velocities, $V_{conv}$ ~10 km/s, the largest convective cell size is $l_{conv} \approx 0.05 R_{star}$, which is comparable to the pressure scale height at $T_0$=30,000K. The upper bound for the turbulent turnover time of a convective cell is $\tau = \frac{l_{conv}}{V_{conv}} \sim 1 \, \text{day}$. Observations of giants suggest that the frequencies of stellar pulsations in late-type luminous stars is on the order of a few to few tens of days. One possible mechanism of Alfvén wave excitation at the wind base is due to acoustic shock waves that dissipate and heat the stellar chromosphere (Judge and Cunz 1993). Another mechanism of Alfvén wave excitation suggests that has been suggested is that they are parametrically amplified at the half frequency of sound waves, due to a swing wave-wave mechanism in the gravitationally stratified stellar atmosphere (Zaqarashvili & Roberts 2002; Zaqarashvili & Belvedere 2005). The expected frequencies of generated acoustic waves in red giants can also be inferred directly from observations of intensity variations of chromospheric emission lines. IUE observations of red giants suggest that the time scale of variations in Mg II h&k emission line is ~ 8.5-12h (Cuntz et al. 1996). Much shorter time scales of emission line variations in Ca II h&k emission lines at ~ 1 hour was obtained by Baliunas et al. (1981) from power spectrum analysis. Therefore, it is reasonable to expect generation of sound waves in giants in a wide range of periods from hours to ~10 days.

## 4. Results of MHD Simulations

We start the simulations by launching Alfvén waves with a broad-band spectrum from the wind base into the initially hydrostatic spherically symmetric atmosphere with a diverging radial magnetic field. The time-dependent solution of Eq. (1-4) is then integrated to a steady-state solution.

Our simulations in Models A through D show that as linear Alfvén waves with the amplitude of the initial transverse velocity, $\delta V_\phi = 0.03 V_A = 4.14$ km/s, launched at the wind base propagate upward in the gravitationally stratified atmosphere, their amplitude, $\delta V_\phi$, increases with height until they become non-linear. This follows from the conservation the Alfvén wave energy flux, $F = \frac{1}{2}\rho_0 \delta V_\phi^2 V_A$, for upward propagating non-dissipative in the gravitationally stratified atmosphere, that do not suffer reflection. The spatial dependence of the $V_\theta, V_\phi$ and $V_r$ components of velocity and the density in Model D in the $r - \theta$ plane at $t = 192\tau_A$ is shown in Figure 2. It is evident that the steady state wind reaches the terminal velocity, $V_r = 0.34 V_A = 46.9$ km/s at $\sim 10 R_*$. Figure 3 shows the $r - \theta$ profiles of the Ohmic dissipation per unit mass, $h = j^2/\rho$, and the $r, \theta$ and $\phi$ components of the local normalized Alfvén velocity over the computational domain for a steady state wind solution at $t = 192\tau_A$. The left lower panel of Figure 3 shows that the formation of localized regions of enhanced Ohmic heating at distances greater than $10 R_*$ from the wind base as the result of a non-linear coupling of transverse motions into longitudinal motions in Model D. This is consistent with the increase of the Alfvén wave amplitudes as they propagate in the gravitationally stratified stellar atmosphere reaching the maximum at distances greater than $10 R_*$ as shown in Figure 4. The radial profiles of the Ohmic heating integrated over $\theta$ in the simulated cone and time-averaged over the longest period of the Alfvén wave spectrum, $P_1$,

$$\bar{h} = \frac{1}{P_1} \iint \frac{j^2(r,\theta)}{\rho(r,\theta)} d\theta\, dt$$

in Models C and D are present in Figure 5. The figure shows that the Ohmic heating, $\bar{h}$, slowly increases from distances $> 5 R_*$ in Model D ( the model of Alfvén-wave spectrum at $\omega_1 = 18\tau^{-1}$ as non-linear effects become more important, while in Model C with twice smaller $\omega_1$, wave amplitudes become non-linear only at $15R_*$. The heating rate reaches its maximum at $\sim 15R_*$ with the value of, $P_{max} = (0.028 - 0.04)B_0^2 V_{A0}/4\pi = (1.2\text{-}1.7) \times 10^5$ ergs/cm$^2$/s for Models D and C respectively. These values imply the formation of warm winds driven by outward propagating non-linear Alfvén waves. The calculated heating rates in these models are about 2 of orders of magnitude greater than that inferred from the surface fluxes observed from spectral lines forming in the lower regions of the stellar wind (Carpenter et al. 1999). The actual value of the heating rate is proportional to the plasma resistivity, $\eta$, while the numerical value of resistivity depends on the smallest spatial scale used in our simulations, and is $\eta = S^{-1} = 10^{-4}$. The classical Spitzer resistivity in fully ionized plasma is $\eta \sim 10^{-11}$ s for the chromospheric temperatures of red giants. However, stellar wind plasma in outer regions is partially ionized, and the effects of ion-neutral coupling will significantly reduce the plasma conductivity with respect to the Spitzer conductivity, and, therefore, correspondingly reduce the value of S by at least 4 orders of magnitude or to $S = 10^7$ (Khodachenko et al. 2004). In contrast to Model C and D, Models A and B with mostly trapped waves output the Ohmic heating that at least 6 orders of magnitude lower. This suggests that models with fully trapped waves produce much cooler winds compared to the models with freely propagating waves.

We should also note that the energy dissipation observed in our simulations caused by numerical compressive viscosity, while the Alfvén waves are dissipated by Ohmic heating and phase mixing effects because heating is localized also along theta direction can all be important sources of the wind heating that will be addressed in the future thermodynamic stellar wind models.

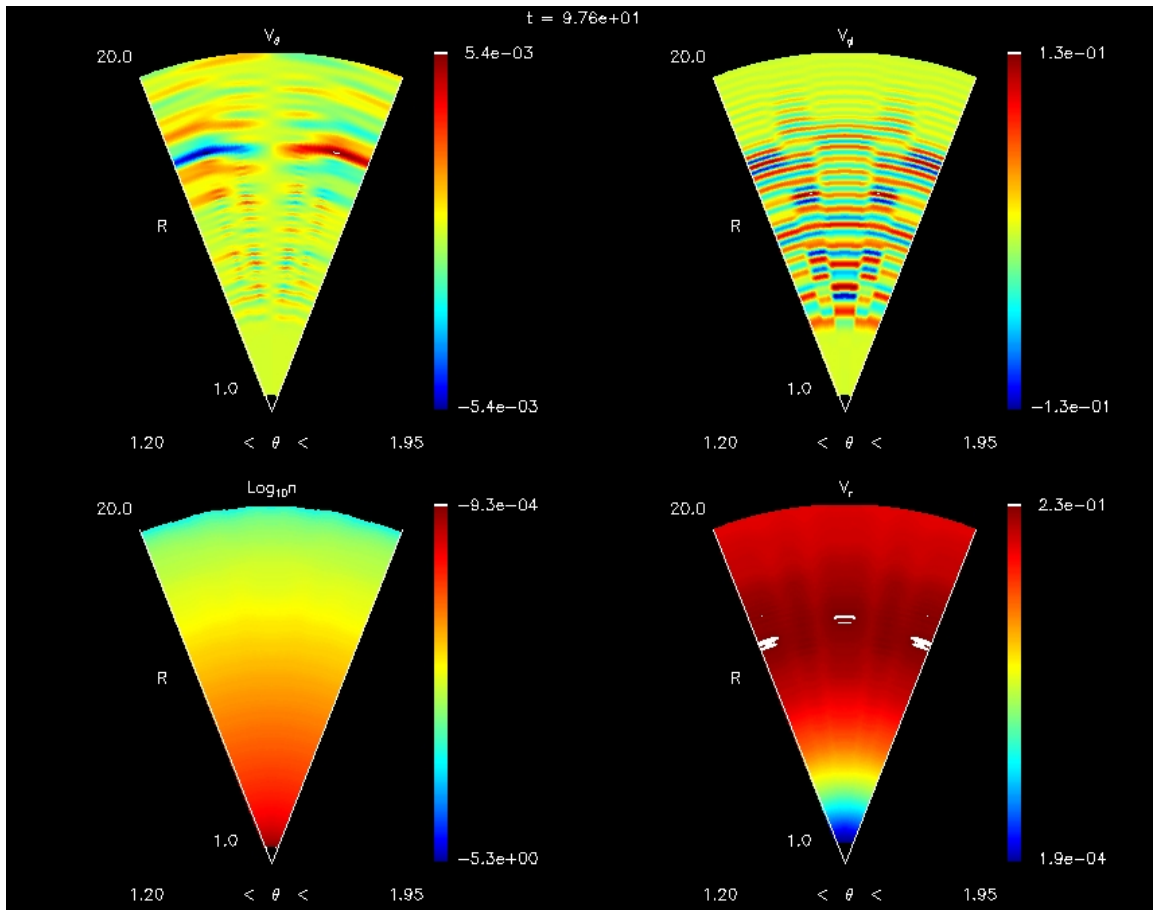

**Figure 2.** The spatial dependence of the $V_\theta, V_\phi$ and $V_r$ components of the wind velocity normalized to the Alfvén velocity at the wind base and the plasma density in Model D in the $r - \theta$ plane at $t = 192\tau_A$

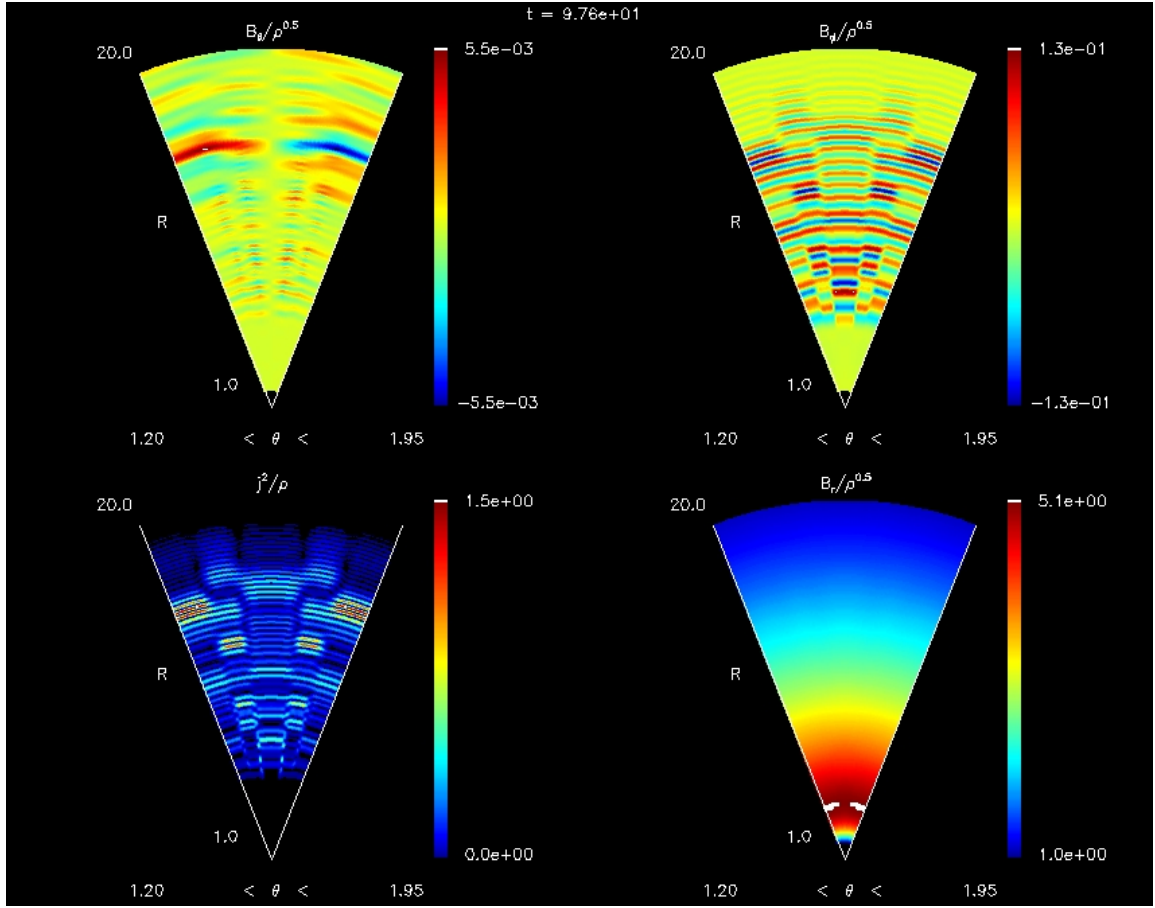

**Figure 3**. The spatial $r - \theta$ dependence of the Alfvén velocity components, $B_\theta / \sqrt{\rho}, B_\varphi / \sqrt{\rho}$ and $B_r / \sqrt{\rho}$, normalized to the Alfvén velocity at the wind base, and the Ohmic heating rate per unit mass, $j^2 / \rho$, normalized to $B_0^2 V_{A0} / 4\pi$ in Model D at $t = 192\tau_A$.

According to Figure 1, the lowest frequency in the wave spectrum in Model B is close to the critical frequency and , therefore, the Alfvén waves are expected to be trapped within a few stellar radii, while waves at all frequencies in Model D are freely propagating. Figure 6 presents the time evolution of the azimuthal velocities in Model B (left panel) and Model D (right panel) as waves propagate through the point at $r = 6.97 R_*$ and $\theta = 90^0$ (the center of the cone) that corresponds to the distance that by

$1R_*$ beyond the sonic point in the stellar atmosphere. It is evident that non-linear Alfvén waves are observed beyond the sonic point as the wind reaches the steady state in Model D, while most of the Alfvén wave flux in Model B the Alfvén waves is trapped below the sonic point well before the wind becomes steady state. The calculation of the total wave flux in Models A and B shows that only less than 0.1% of the total energy penetrates distances greater than $1R_*$ from the wind base.

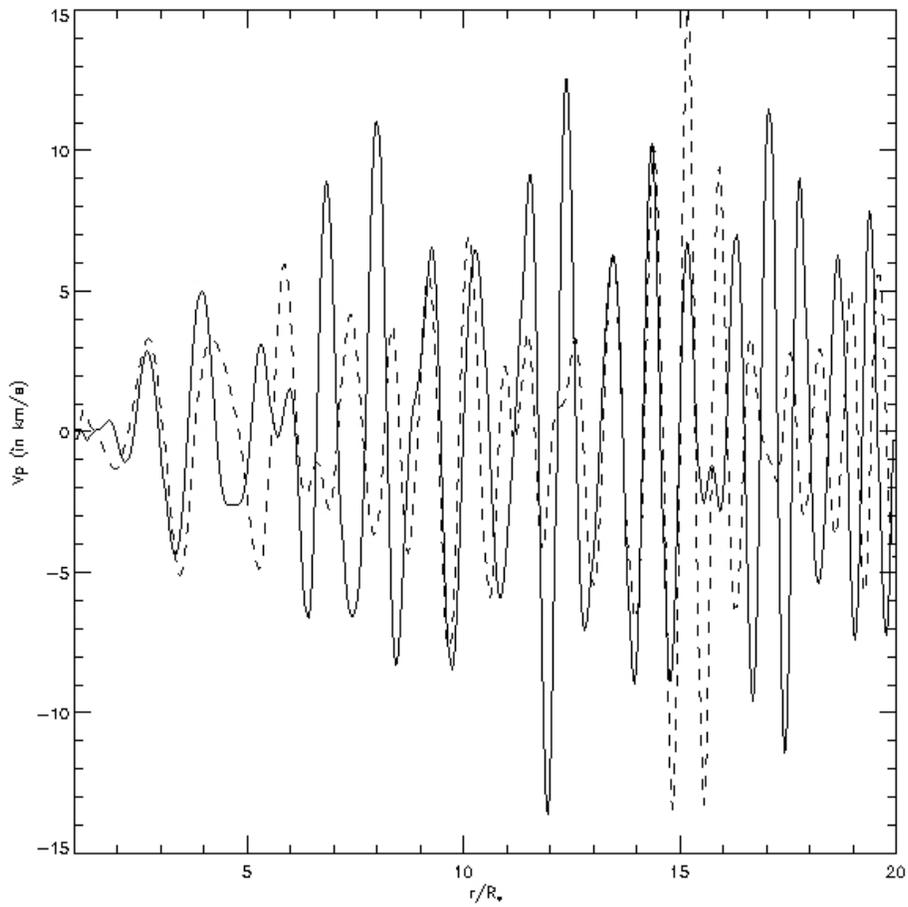

Figure 4. The radial profile of $V_\phi$ components of the steady state wind velocity in Model D at $t = 192\tau_A$ for the center of the simulated cone (solid line) and $\theta = 45^0$ (dashed line).

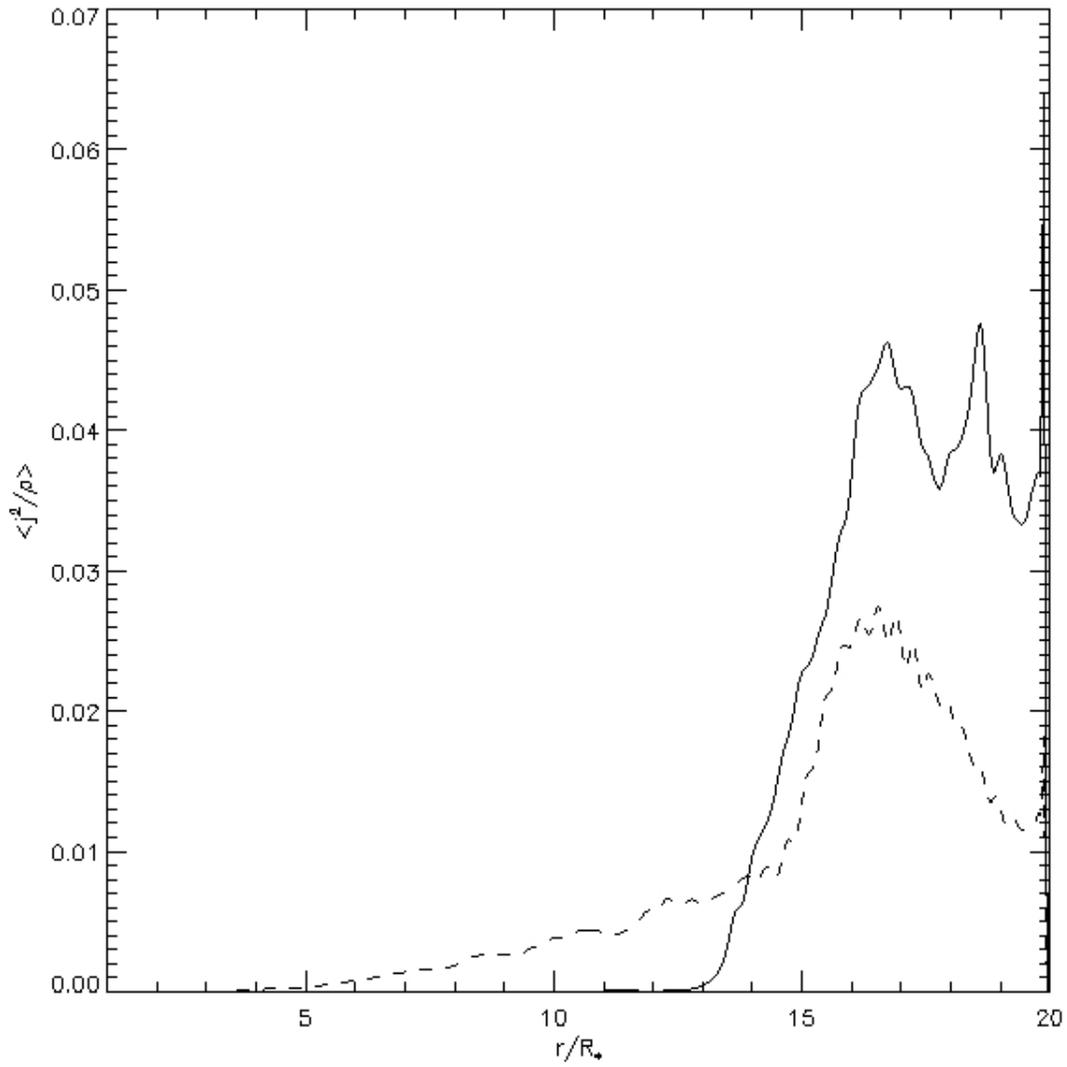

**Figure 5**. The radial profiles of the Ohmic heating per unit mass integrated over $\theta$ in the simulated cone and time-averaged over the longest period of the Alfvén wave spectrum, $< j^2/\rho >$, normalized to $B_0^2 V_{A0}/4\pi$ in the steady state wind Model C (solid line) and in Model D (dashed line) at $\theta = 0^0$.

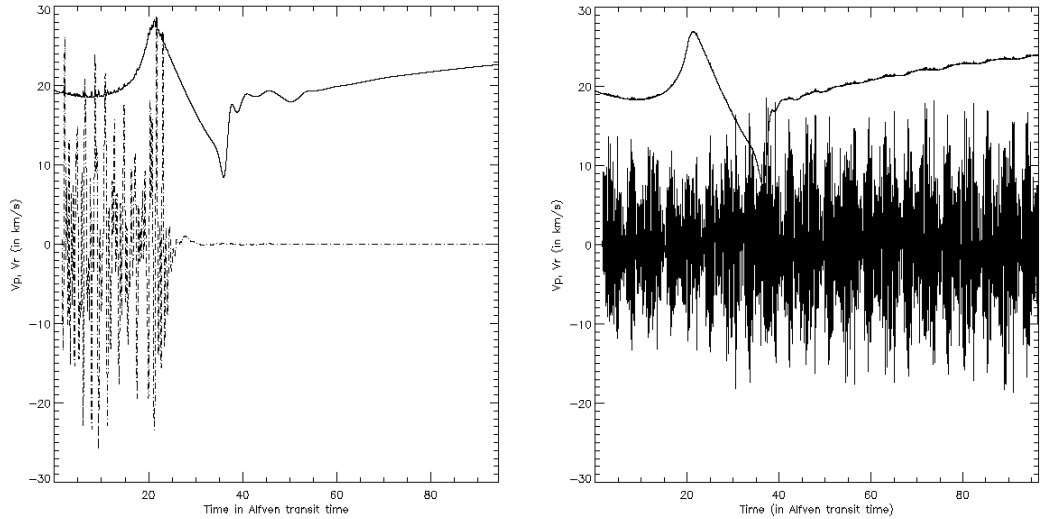

**Figure 6.** The time evolution of the radial ( upper plot) and the azimuthal (lower plot) components of the wind velocity in Model B (left panel, reflected waves) and Model D (right panel, propagating waves) at r=6.97R$_*$ , $\theta = 90^0$.

Figure 7 shows the radial profiles of the wind terminal velocities in Models A, B, C and D. The figure shows that radial acceleration of stellar winds in the first 5 stellar radii is generally insensitive to the lowest frequency of Alfvén waves, while the wind terminal velocity increases by 20% in Model D with respect to Model A. The figure shows that in the models with trapped waves (Model A and B), the wind radial velocity is about 30 km/s at , while the wind terminal velocity ~42-45 km/s at . The Models C and D outputs greater wind velocities reaching over 50 km/s at . The calculated velocities at in Models A and B are consistent with the terminal wind velocity for α Tau like star (see Table 1, Robinson et al. 1998).

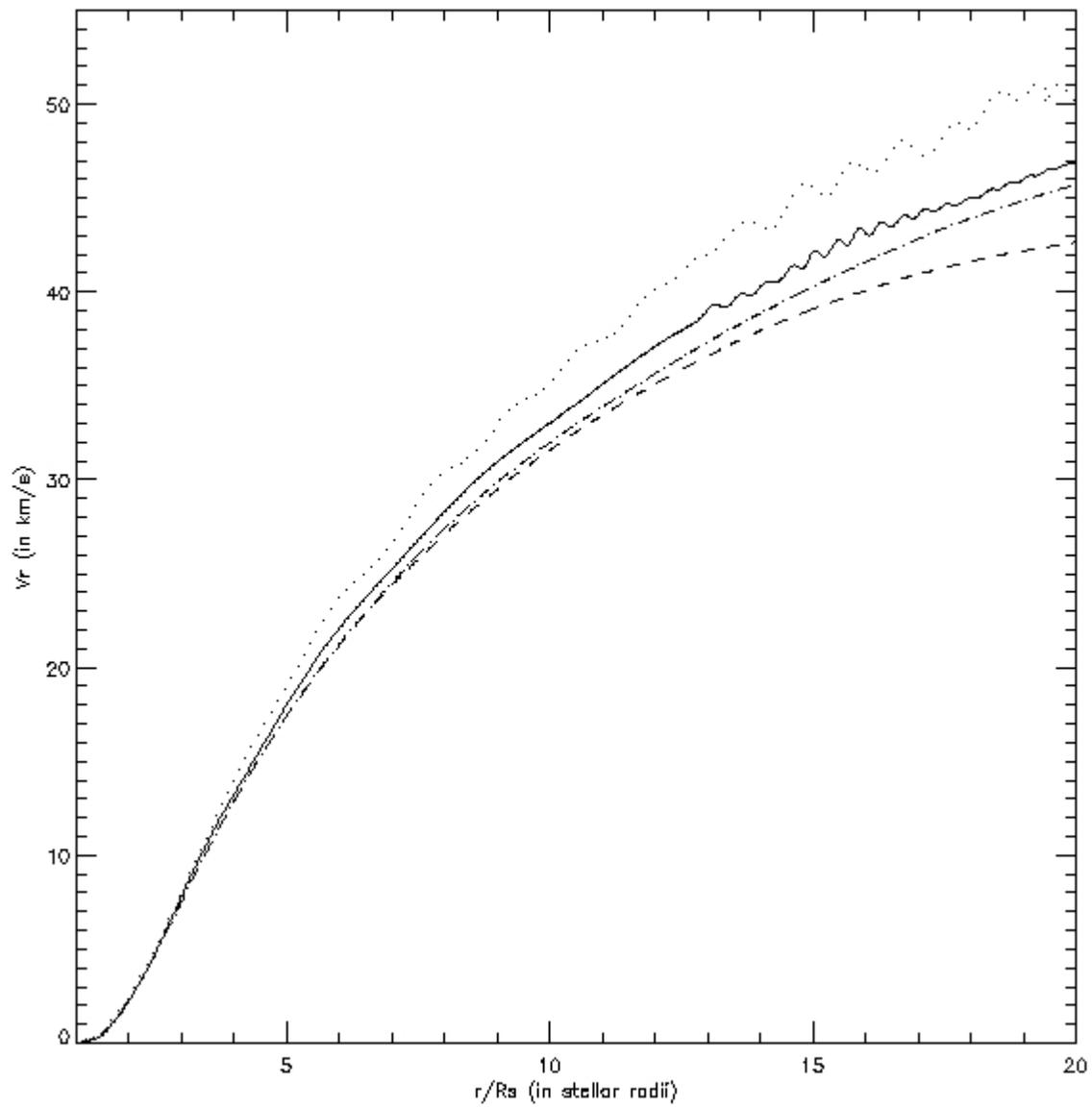

**Figure 7.** The radial profiles of radial velocities of the steady state winds in Model A (solid), Model B (dash-dot), Model C (dot) and Model D (dash-double dot).

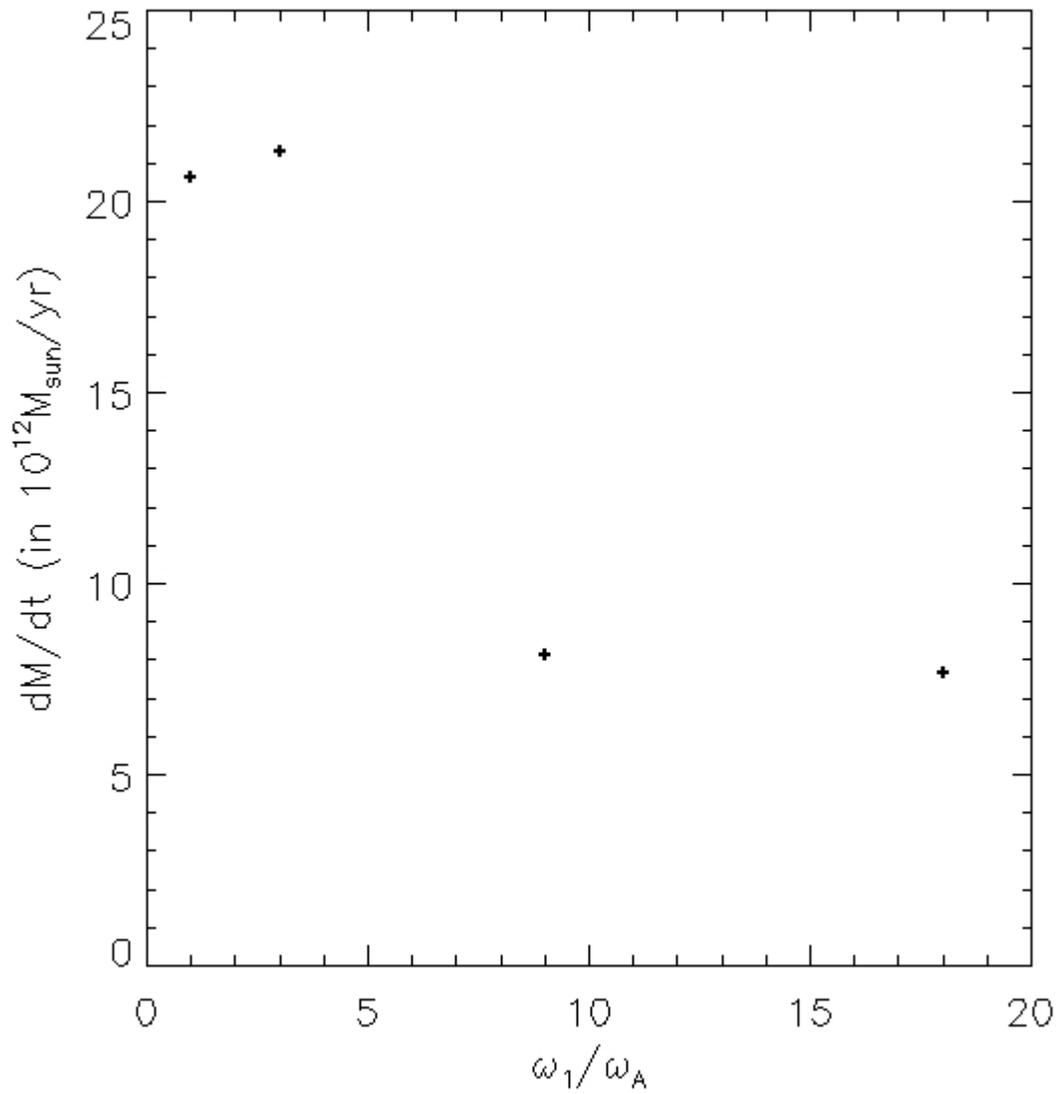

Figure 8. The total mass loss rate as a function of the lowest frequency in the broad-band Alfven wave spectrum (Model A, B, C and D). The figure shows the sharp increase in the mass loss rate in the model with mostly trapped waves.

The time-averaged, normalized mass flux per unit surface area of the steady state wind is

$$\dot{M}_0 = \frac{1}{\Delta t} \int_{t1}^{t2} \rho(t) V_r(t) dt \qquad (10)$$

where integration of the mass flux is performed over 10 wave periods. The Figure 8 shows the total mass flux, $\dot{M}_0$, as a function of the lowest frequency in four simulated broad-band spectra (Models A through D). The figure shows that as waves get reflected at lower regions in the stellar atmosphere where density is greater, they drive more massive winds. Summarizing our findings from Figures 7 and 8, we conclude that our simulations output faster and more tenuous (lower mass-loss rate) winds as the lowest frequency in the spectrum of waves increases and waves become less trapped in the stellar atmosphere. Our results derived from fully non-linear simulations, therefore, are consistent with earlier studies (e.g. Holzer & McGregor 1985).

The normalized mass flux per unit surface area, $\dot{M}_0$, in Model A ( the wind model driven by fully trapped Alfvén waves) is a factor of ~3 greater than that calculated in Model D (the model of freely propagating waves). The total mass loss rate from the stellar surface in the star filled by an open magnetic field is then given as

$$\dot{M} = \dot{M}_0 \sqrt{\frac{\rho_0}{4\pi}} \Phi_0, \qquad (11)$$

where, $\Phi_0 = 4\pi f B_0 R_*^2$, is the total magnetic flux of the region(s) in the star filled with the open magnetic field with the filling factor $f$ and the magnetic field, $B_0$, and $\rho_0$ is the plasma density at the wind base. Comparison of the calculated mass loss rate from Equation (11) with the one observed in α Tau, $\dot{M}_{obs} = 1.6 \times 10^{-11} M_{sun}/yr$ (see Table 1), yields the average value of the open magnetic field over the stellar surface, $fB_0 \sim 0.75$, in Model D. For the assumed value of the radial magnetic field of 2Gs, $f \sim 0.37$, that implies an anisotropic wind from a cool giant star like α Tau. Thus, we conclude that Model A and Model B output the wind velocities and the mass loss rates consistent with the observed parameters for an α Tau-like red giant.

Moreover, Equation (11) also suggests that the mass loss rate may vary on timescales of the emergence and evolution of the magnetic flux associated with an open magnetic field or on timescale of the rotation period of the star that reveals changes in distribution of an open magnetic flux of the active regions in the stellar atmosphere. While the timescale of magnetic field amplification by giant cell convection is in the order of 25 years (Dorch 2004), the rotation periods of a typical giant (like α Tau) is on the order of 1 year. This suggests that asymmetries in the filling factors of open magnetic filed in two hemispheres can explain variations of the measured mass loss rate as $\sim \Phi_0$, as the star rotates about its axis. A possible indication on the role of variations in the filling with stellar rotation comes from HST/GHRS observations of Fe II and CII] lines forming in the wind of α Tau that have been taken in 1990, 1994 and 1995 by Robinson et al. (1998). For instance, while the line profile CII] line at three epochs did not change, the integrated line fluxes in 1994 and 1996 increased by 10%-30% with respect to 1990 with variation of the UV continuum (2320-2360Å) emission in 1994 to 1996 by a factor of 2 and 1.77 respectively. Plasma density derived from the optically thin and density sensitive ratios of the CII] lines at three epochs shows no change at three epochs. If we assume that variations in line fluxes observed at three epochs were caused by variation in heating rates due to Alfvén wave dissipation, then the total line flux over the stellar surface, A, is

$$F_t = \frac{1}{2} \int_A \rho_0 <\delta V_\phi^2> V_A dA \propto \rho_0 <\delta V_\phi^2> \Phi_0 \qquad (12)$$

Equation (12) then suggests that increase in the line fluxes is due to the increased magnetic flux at three observed epochs by 10%-30% in 1994 and 1996 relative to 1990 because the density and the average amplitude of unresolved wave motions that contributes to the non-thermal broadening remain the same. The variation of magnetic flux with stellar rotation and/or stellar magnetic cycle that follows from Eq. (11) also can explain mass loss rate variations in λ Vel observed at 3 epochs, a factor of 6 variation in the mass loss rates over 18 months in K757 (M15) and by a factor of 2 in K341 (M15) and L72 (M13) (Mullan et al. 1998; Mészáros et al. 2009).

## 4. Conclusions

In the current study we simulated slow and massive winds from evolved giants driven by non-linear broadband Alfvén waves propagating in the gravitationally stratified stellar atmosphere of a red giant with stellar parameters of α Tau. The simulations were performed with our 2.5D time-dependent fully nonlinear MHD code for four time-dependent wind models characterized by the lowest frequency in the broadband spectrum of Alfvén-wave driven winds. We found that fully reflected non-linear Alfvén waves deposit energy and momentum well below the sonic point ( Model A & B) and produce ~3 times more massive and 20% slower wind than that formed by the fully propagating Alfvén waves (Model C & D). This result is consistent with solar wind studies and our previous studies of stellar winds (Ofman and Davila 1998; Airapetian et al. 2000) and, for the first time, provides the estimate of the contribution of trapped Alfvén waves to the mass-loss rates from red giants. The calculated mass-loss rates and terminal velocities consistently reproduce the observed wind velocity and mass-loss rate from cool giants similar to α Tau. Specifically, when the simulated total mass loss rate is compared with the measured mass loss rate in α Tau, we conclude that the $_fB$ factor, the average value of the open magnetic field over the stellar surface, points at an anisotropic wind from the star with the filling factor ~0.37. In addition, our simulations show that the mass loss rates from Alfvén wave driven winds depend on the total magnetic flux in open magnetic field in the stellar atmosphere that can be either rotationally modulated or produced by dynamo motion and transport to the stellar surface. Therefore, our model predicts that the stellar winds driven by non-linear Alfvén waves from red giants are expected to be anisotropic in space and variable in time on a timescale of stellar rotation and stellar cycle. The results obtained in this paper are also consistent with detections of anisotropic mass-loss from M giants using diffraction-limited IR imaging at Keck (Tuthill et al. 1998; Hofmann et al. 2000) and observations of time varying mass loss rates from a number of late-type giants discussed in the paper.

The current simulations represent dynamic properties of 2D nearly isothermal winds. However, in the near future, we are planning to include the effects of Alfvén wave dissipation in the energy equation due to combination of Ohmic, and viscous heating mechanisms and wave phase mixing. Thus, the addition of the energy equation with improved equation of state will be capable of reproducing not only dynamics but also thermodynamic parameters of winds from red giant stars.

**Acknowledgments.** VA was supported by NASA grant NNX10AK22G. LO was supported by NASA grants NNX08AF85G, NNX10AC56G, NNX08AV88G.